\documentclass[conference]{IEEEtran}
\IEEEoverridecommandlockouts

\usepackage{cite}
\usepackage{amsmath,amssymb,amsfonts}
\usepackage{algorithmic}
\usepackage{graphicx}
\usepackage{textcomp}
\usepackage{xcolor}
\usepackage{multirow}
\usepackage{url}
\usepackage{diagbox}

\begin{document}

\title{Proximity Neighbor Selection \\in Blockchain Networks}

\newif\ifblind
\blindfalse

\ifblind
\else
\author{
\IEEEauthorblockN{Yusuke Aoki%
, and Kazuyuki Shudo%
}
 \IEEEauthorblockA{Tokyo Institute of Technology \\
2-12-1 Ookayama, Meguro, Tokyo, Japan \\
}}
\fi

\maketitle

\begin{abstract}
Blockchains have attracted a great deal of attention as a technology for the distributed management of register information at multiple nodes without a centralized system. However, they possess the drawbacks of low transaction throughput and long approval time. These problems can be addressed by shortening the block generation interval; however, shortening this interval alone has the effect of increasing the frequency of forks. In this study, we aim to shorten the block generation interval without increasing the fork generation rate by improving the network topology of the nodes and shortening the propagation time.
We propose a neighbor node selection method forming a network topology with a short block propagation time. A blockchain simulator is used to demonstrate the effect of the proposed neighbor node selection method on the propagation delay of the network.
This result indicates that the proposed method improves block propagation time.
\end{abstract}


\section{Introduction}

Blockchain is a distributed ledger technology that appeared as the core technology of the cryptocurrency Bitcoin \cite{nakamoto2008bitcoin}, which is developed by the name of Satoshi Nakamoto.
In recent years, many cryptocurrencies using blockchain have been ledger developed and operated, and blockchains have attracted increasing attention.
Blockchains can manage information securely and protect it from tampering, even if multiple malicious nodes are present.
Additionally, no central management is required, and a blockchain system can operate independently.
These features have proven to be very useful, and their application is being studied not only in cryptocurrency but in a wide range of fields.

Although blockchains have many advantages, they also have several drawbacks.
The primary problems involve the low throughput of transactions and the approval time for a transaction.
These obstacles can be overcome by shortening the block generation interval.
However, if the block generation interval alone is shortened, the block propagation time will not be sufficiently shorter than the block generation interval; consequently, the frequency of forks will increase, leading to increased security risk \cite{sompolinsky2015}.

In this study, we construct a network topology with a short block propagation time as a policy for overcoming the aforementioned challenges.
Because a blockchain network is a peer-to-peer network without a central management system, the network topology is determined by each node's selection of neighbor nodes.
Therefore, in this paper, we propose a neighbor node selection algorithm for forming a network topology with a short block propagation time.
In the proposed algorithm, each node evaluates other nodes using a score based on the speed of block delivery, and preferentially selects a node with a good score to be a neighbor node.
An improved block propagation time can be achieved if each node dynamically changes neighbor nodes based on information that can be obtained during blockchain operation.

Using a simulator, we demonstrated that the proposed neighbor node selection algorithm improves block propagation time.

The remainder of this paper is structured as follows. Section \ref{sec:blockchain} provides background information on blockchains, while Section \ref{sec:proposed} proposes a neighbor node selection algorithm to improve transaction throughput.
Section \ref{sec:experiment} describes experiments performed using a simulator to confirm the proposed method.
Section \ref{sec:summary} presents conclusions and ideas for future work.

\begin{figure}[t]
  \begin{center}
    \includegraphics[width=8.35cm]{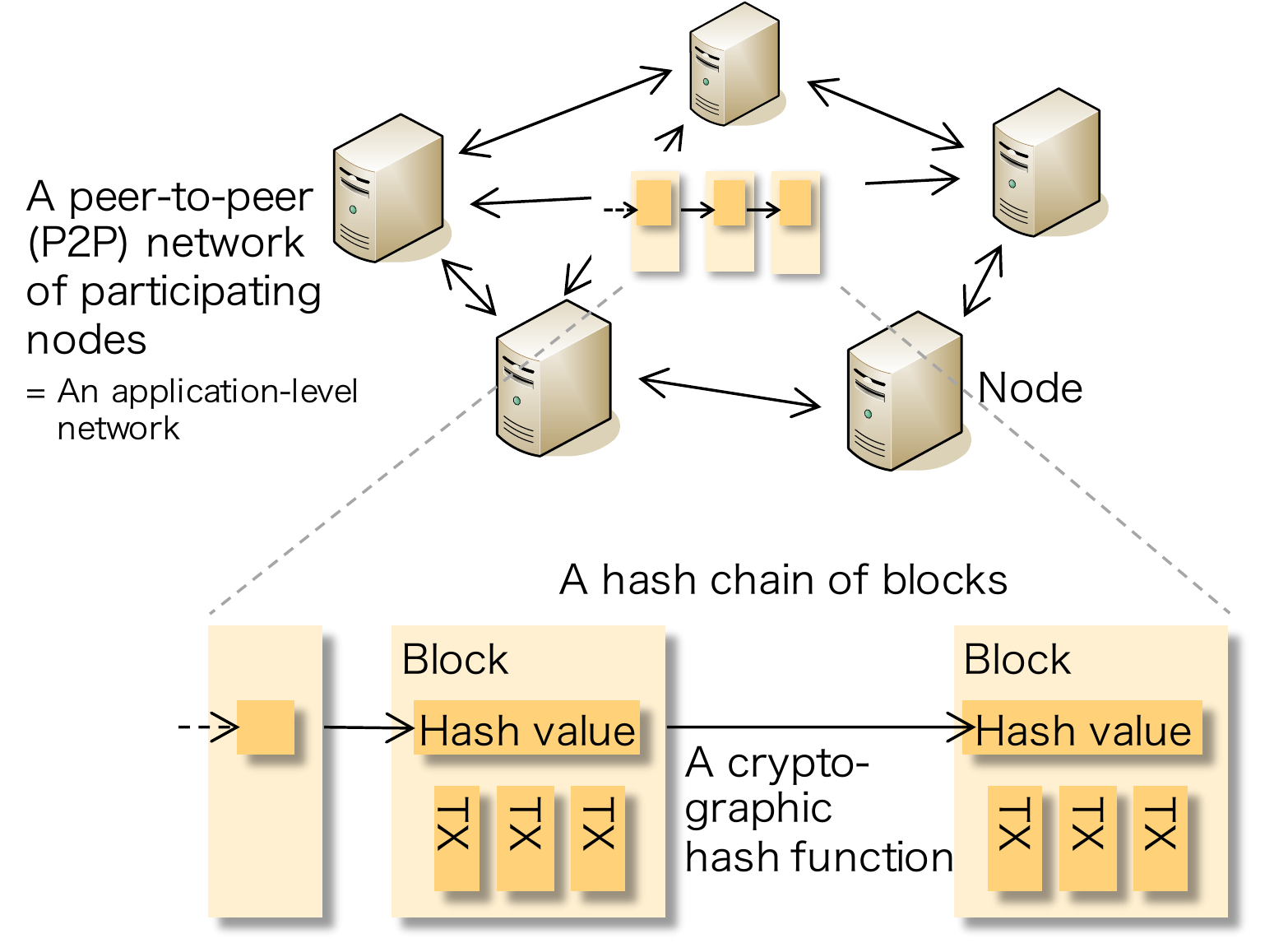}
    \caption{Blockchain overview.}
    \label{fig:block}
  \end{center}
\vspace{-3ex}
\end{figure}

\section{Blockchain}
\label{sec:blockchain}

Blockchain is a distributed ledger technology regarded as part of Bitcoin, proposed by Satoshi Nakamoto.
The nodes of a blockchain constitute a peer-to-peer network.
Because blockchains have no central management, a consensus algorithm has been devised to ensure a consistent ledger without contradiction between nodes.
Most blockchain consensus algorithms are Byzantine fault-tolerance \cite{lamport1982byzantine} and have the useful feature that even if a malicious node intentionally propagates false information, the blockchain can form a correct consensus.
\subsection{Blockchain technology}
\subsubsection{Transaction propagation}
Data to be recorded in blockchains are called transaction, and they are broadcast to the network through nodes participating in the blockchain network.
The broadcasted transactions are stored in the transaction pool of each node; however, they are not approved yet and are not recorded in the ledger.

\subsubsection{Consensus}

Transactions are only approved by being included in a block.
A block contains multiple transactions, including the hash value of the immediately preceding block, and a freely settable variable called nonce.
As illustrated in Fig. \ref{fig:block}, the hash value of the previous block is included in each block; therefore, all blocks are connected in a chain referred to a blockchain.
This series of blocks is a ledger that records transactions.

Each node generates a block that includes transactions in its transaction pool.
The generated block is propagated through the network and accepted and shared by other nodes.
At this point, the transaction is accepted for the first time and recorded in the ledger.

At this time, it is important that the method to decide which node generates the block.
Because blockchains lack have a central management mechanism, it is necessary to autonomously select a node to generate a block.
Furthermore, even in a blockchain network in which multiple malicious nodes participate, it is necessary to design so that the malicious nodes can not  falsify the ledger data by monopolizing the block generation right.
Several algorithms have been proposed for deciding which nodes generate a block.
The Proof of Work (PoW) system used in Bitcoin is one of the most widely known algorithms.

\subsubsection{Proof of Work}

In PoW, a node for generating a new block is selected based on its computing power.
Each block includes a value that is freely set by each node called a nonce, and each node locates a block whose hash value is below a certain threshold while changing this nonce.
Only blocks below the threshold are regarded as formal blocks.
Therefore, a node that discovers a nonce satisfying the aforementioned condition can generate a new block.
The difficulty of block generation can be adjusted by changing the threshold value.
The process of calculating the hash value of an entire block while changing the nonce is called mining.
In PoW, each node can generate a new block with a probability proportional to its computing power.

\subsubsection{Network}

The nodes involved in a blockchain form a peer-to-peer network in which
transactions and blocks are broadcasted.

A blockchain peer-to-peer network is an unstructured network, and there are no rules for structurally defining connections. 
When a node creates a new connection, it selects a destination node from the list of participating nodes loosely shared throughout the network.
Nodes participating in the network exchange information regarding nodes that are known to them.
When a new neighbor node is required, a node selects one using this node information.

In Bitcoin's reference implementation, Bitcoin Core \cite{Bitcoincore}, new connections are generated on limited occasions, such as when a node joins the network or when an existing neighbor node is disconnected.
Therefore, the topology of Bitcoin's network does not change significantly in a short period of time \cite{miller2015discovering}.
A connection that a node actively decides the destination and creates is called Outbound, while a connection created in response to other node's Outbound is called Inbound.
Both connections send and receive blocks and transactions in the same way.
By default, eight Outbounds are created, and no more than 125 Inbounds are created.

\begin{figure}[t]
 \begin{center}
    \includegraphics[width=0.23\textwidth]{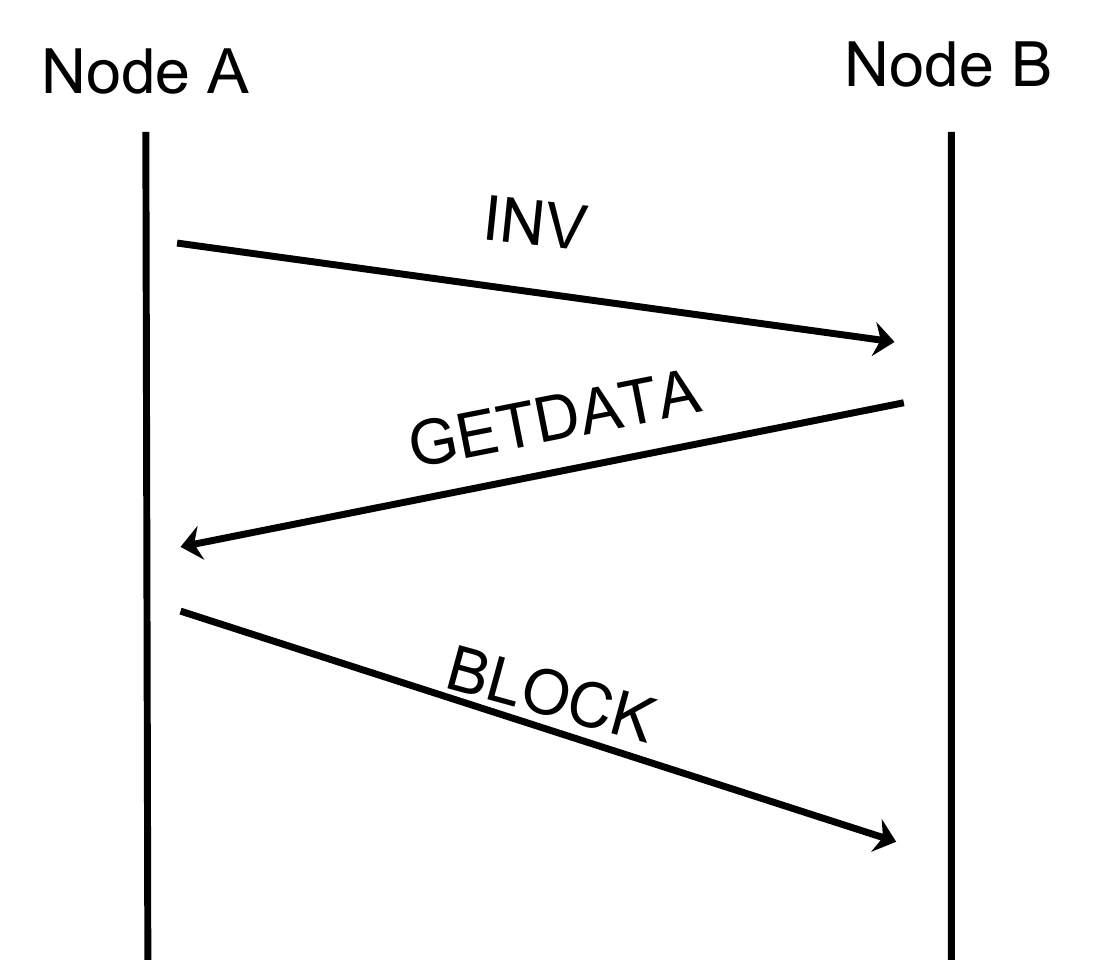}
    \caption{Block propagation from node A to node B.}
    \label{fig:inv}
  \end{center}
\vspace{-3ex}
\end{figure}

In simple protocols, block transmission/reception takes place using the protocol presented in Fig. \ref{fig:inv}.
Before sending a block, a node sends an INV message and verifies whether the destination node already possesses the block.
If the node receiving the INV message does not possess the block, it responds with a GETDATA message and waits for block reception.
By using this protocol, the unnecessary transmission of blocks containing a large amount of data is prevented.
 
\subsubsection{Fork}

If the next block is generated before one block propagates the entire network, tow different blocks get propagated through the network, which is called a fork.
A fork result in each node possessing a different block as the latest block, which leads to a loss of data consistency.
Additionally,  a high incidence of forks often allows attacks such as selfish mining \cite{eyal2018majority}, and it is a large security risk.
To prevent forks from appearing in existing blockchains, the difficulty of block generation is increased, and the block generation interval is lengthened to ensure that multiple blocks are not generated simultaneously. 
In the case of Bitcoin, the difficulty of block generation is adjusted to allow one block to be generated every 10 min.

\subsection{Transaction approval speed}

Two challenges faced by blockchains are the interval between the occurrence of a transaction and the approval is long and that the throughput of the transaction is low \cite{croman2016scaling}.

\subsubsection{Long approval interval}

Transactions are not approved until they are included in a block.
Therefore, they are not approved immediately upon being issued, but only after the following block is generated.
For Bitcoin, the block generation interval is 10 minutes; therefore, on average, it takes approximately 5 minutes for a transaction be approved. Additionally, the approval interval experiences a further increase when considering the possibility that forks occur and blocks become invalid.
In most cases, if approximately six blocks are generated as a subsequent block, it is considered that the block will not be invalidated.
Therefore, the approval of a transaction may take up to an hour or more.

A method for reducing the time required for approving a transaction involves shortening the block generation interval. there is a method to shorten the block generation interval.
However, simply shortening this interval increases the incidence of forks.

\subsubsection{Low throughput}

The throughput of a blockchain is obtained by dividing the number of transactions in a block by the block generation interval.
For Bitcoin, the upper limit of the number of transactions in one block is approximately 4,000, and the block generation interval is 10 min (600 s). Therefore, the upper limit of the throughput is approximately seven transactions per second.
This is a very small value compared to the average throughput of Visa, which is approximately 1,700 transactions per second \cite{Visa1700tps}, or the average throughput of PayPal, which is approximately 350 transactions per second \cite{Paypal2019Q1}.

There are two methods for improving the throughput: increasing the number of transactions in one block and shortening the block generation interval.
Using the former method, as the block size increases, the block propagation time between nodes also increases\cite{decker2013information} and it becomes necessary to increase the block generation interval to suppress the occurrence of forks.
Consequently, the throughput does not improve.
Using the latter method, shortening the block generation interval is also problematic because it increases the incidence of forks.

\subsection{Reducing block propagation time}

Although the long approval interval and low throughout can both be addressed by shortening the block generation interval a problem occurs whereby the incidence of forks increases.
To shorten the block generation interval without increasing the incidence of forks, it is necessary to simultaneously shorten the block propagation time.
This, in turn, influences the scale of the number of nodes in a blockchain network.
As the number of nodes in the network increases, the block propagation time through the entire network becomes longer, and the incidence of forks increases.
To safely scale the number of nodes, it is necessary to reduce the block propagation time.

In this paper, we attempt to improve the efficiency of the network topology by shortening the block propagation time.
Because blockchain networks are peer-to-peer networks without a central administrator, the topology of the network is dependent on the manner in which neighbor nodes are selected.
We aim to decrease the propagation time of the network by improving the neighbor selection algorithm.

\section{Proposed Algorithm}
\label{sec:proposed}

Proximity neighbor selection \cite{miyao2013proximity,Gummadi2003proximity} is a proven technique to improve propagation performance in peer-to-peer networks.
This section describes our proposed algorithm to to perform proximity neighbor selection in blockchain networks.
In the proposed algorithm, we designed each node to connect with the node that sent the INV message earlier to it.
It is possible to construct a network topology in which nodes with good network conditions such as bandwidth can efficiently deliver blocks.

\subsection{Node scoring}
\label{sec:score}
In the proposed algorithm, each node rates the node that sent it an INV message and determines the connection priority.
The score is based on the amount of time between block generation and the receipt of the INV message.
Each time a node receives an INV message from another node, it updates that node's score.

In (1), $SCORE_N$ is the score of node $N$ when an INV message from that node is received. 
The block creation time is denoted $T_{Block}$, and the arrival time of the INV message is denoted $T_{INV}$.
If INV message of any block has not been received from node $N$ until then, the score is updated as follows:
\begin{equation}
SCORE_N \Leftarrow T_{INV} - T_{Block}
\end{equation}
However, If INV message of any block has been received from node $N$ until then, the score is updated as follows. 
\begin{equation}
\label{score}
SCORE_N \Leftarrow (1 - P) \times SCORE_N +  P \times ( T_{INV} - T_{Block} )
\end{equation}
Here, $P$ is a weighting parameter in the range of $[0 \ 1]$.
The appropriate value of $P$ is determined experimentally in Section \ref{sec:experiment}.

In a regular blockchain, blocks contain the block generation time information, and each node can obtain the reception time of an INV message.
Therefore, additional information is not required to calculate the node score. 

\subsection{Update neighbor nodes}
Each node reselects all of its neighbor nodes at regular intervals using the aforementioned node score.
In this paper, neighbor nodes are reselected whenever 10 blocks are received.
Each node selects new neighbor nodes in order of increasing scores.

However, $K$ neighbor nodes are randomly selected from all nodes in the network to obtain information on the new node.
In many blockchains, information on nodes in the network is regularly propagated through the network, making it possible to select nodes from the entire network at random.
We conducted an experiment that involved changing the value of parameter $K$ in order to determine the appropriate parameters; this is described in more detail in Section IV.

The connection created by the neighbor node selection is an Outbound connection, which is an active connection.
To prevent the concentration of Outbound connections on only several nodes, the number of Inbound connections generated corresponding to  The Outbound connection of other nodes is set to 30 or less.

\begin{table}[t]
\begin{center}
	\caption{Parameters of simulator proposed by Gervais et al.}
	\label{table:actualdata}
	\begin{tabular} {lccc} \hline
		 Parameter	& Bitcoin 	& Litecoin	&Dogecoin\\ \hline
		\# of the nodes	&6000 	 &800    &600\\
		Block interval	&10 min	&2 min 30 sec	&1 min\\
		Block size	&534 KiB	&6.11 KiB	&8 KiB\\
		\# of connections &\multicolumn{3}{c}{
Distribution according to Miller et al.  \cite{miller2015discovering}}\\
		Geographical distribution	&\multicolumn{3}{c}{Distribution according to actual blockchains}\\
		Bandwidth	 & \multicolumn{3}{c}{\multirow{2}{*}{6 regional bandwidth and propagation delay}}\\
		propagation delay &\\ \hline
	\end{tabular}
	\end{center}
\end{table}

\subsection{Impact on security}

Because our proposed algorithm imparts regularity on the blockchain network topology, close attention must be paid to attacks such as the Eclipse attack\cite{heilman2015eclipse}, in which malicious nodes make use of topology.
The Eclipse attack is an attack that occupies replaces all of the neighbor nodes of a target node or a target group of nodes with malicious nodes, thereby separating the target from the network.
Lowering the randomness of neighbor node selection thus increases the risk of the malicious manipulation of neighbor nodes.
However, in our proposed algorithm, the node that sends the block faster is selected as the neighbor node. In other words, the node that wins the block distribution competition between nodes is preferentially selected as the neighbor node.
Therefore, because maliciously occupying neighbor nodes involves winning a competition between other nodes, the cost of the attack increases.
Additionally, an existing solution\cite{heilman2015eclipse} that evaluates individual connections such as the Ban system and Feeler connection, can be applied to the connections determined by the proposed algorithm.
Furthermore, because $K$ nodes are randomly selected as neighbor nodes, many of the same measures used in regular blockchains can be used for these neighbor nodes.

\section{Experiment}
\label{sec:experiment}

The proposed algorithm is evaluated using a blockchain simulator developed by Aoki et al. \cite{SimBlock-CryBlock,SimBlock-ICBC}.
In all the experiments, the unlisted parts of parameters simulate the Bitcoin environment in 2015 examined by Gervais et al. \cite{gervais2016security}.
The parameters in Table \ref{table:actualdata} were reproduced in a similar manner as in the simulator proposed by Gervais et al.

\begin{figure}[t]
 \begin{center}
    \includegraphics[width=8.35cm]{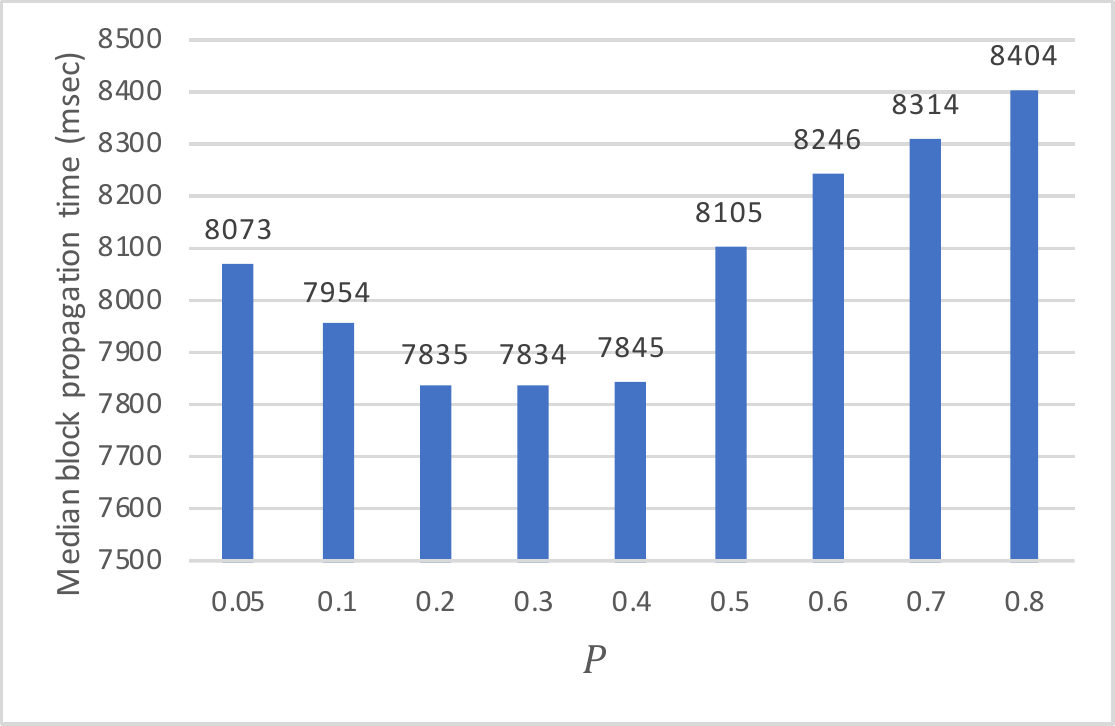} 
    \caption{Median block propagation time for different values of parameter $P$.}
    \label{fig:presult}
  \end{center}
\vspace{-3ex}
\end{figure}

\begin{figure}[t]
 \begin{center}
    \includegraphics[width=8.35cm]{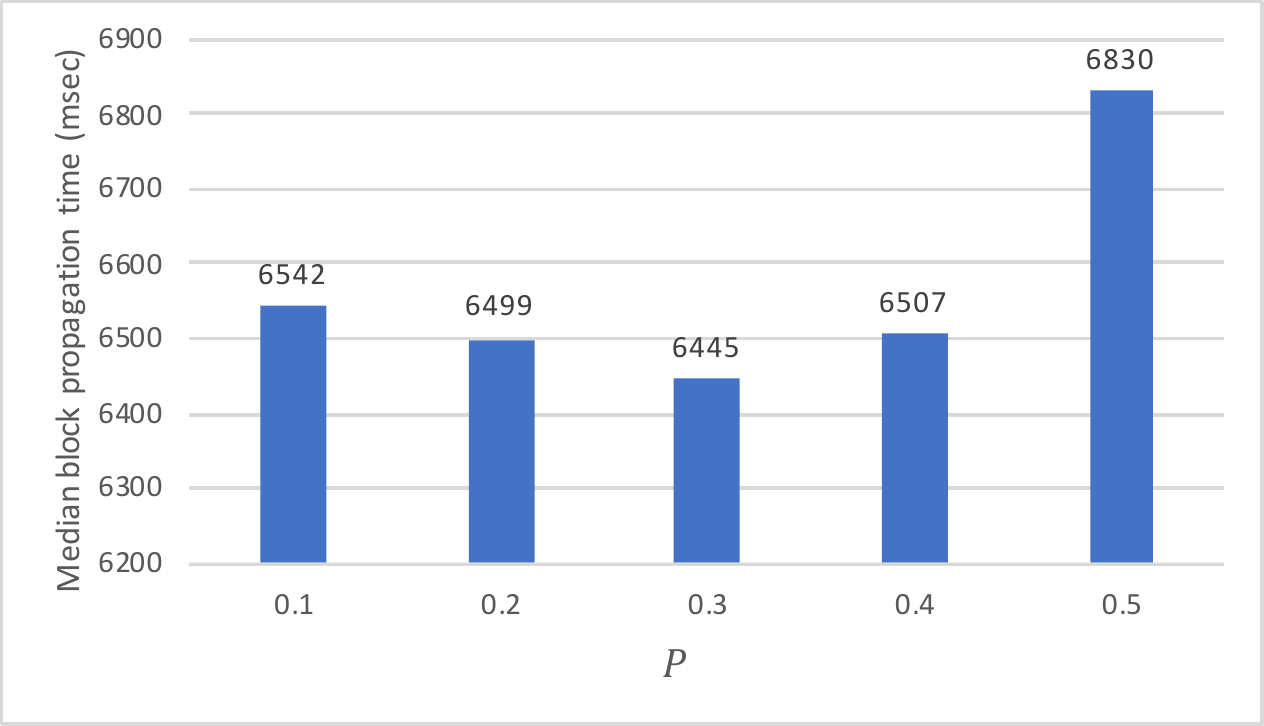} 
    \caption{Median block propagation time for different values of parameter $P$ (1,000 nodes).}
    \label{fig:presult1000}
  \end{center}
\vspace{-3ex}
\end{figure}

\subsection{Preliminary experiment}
In the proposed algorithm, there are two parameters, the weight of score $P$ and the number of neighbor nodes $K$ selected randomly.
We performed a preliminary experiment to determine the appropriate values for these two parameters.

\subsubsection{Weighting parameter $P$}
In the proposed algorithm, upon each receipt of an INV message, the node score is updated using (\ref{score}), defined in Section \ref{sec:score}.
In this experiment, we changed the weighting parameter $P$ and determined the appropriate $P$ value.

We performed the simulation until 5,000 blocks were generated; we then measured the median value of the block propagation time, which is the average value of all 5000 blocks.
In this experiment, $K = 1,$.
Fig. \ref{fig:presult} presents the median block propagation times for different parameters $P$. 
It can be seen that the median block propagation time is shortest (7.83 seconds) when $P = 0.3$

Next, we confirmed that the optimal parameters did not change depending on the number of nodes participating in the network.
Fig. \ref{fig:presult1000} presents the experimental results when the number of participating nodes was changed to 1,000.

The results confirmed the same tendency for 1,000 nodes as for 6,000 nodes.

\subsubsection{Number of randomly selected node $K$}
In the proposed algorithm, $K$ nodes are randomly selected as neighbor nodes from the entire network.
In this experiment, we changed parameter $K$ and compared the median block propagation times to identify an appropriate $K$.

\begin{figure}[t]
 \begin{center}
    \includegraphics[width=8.35cm]{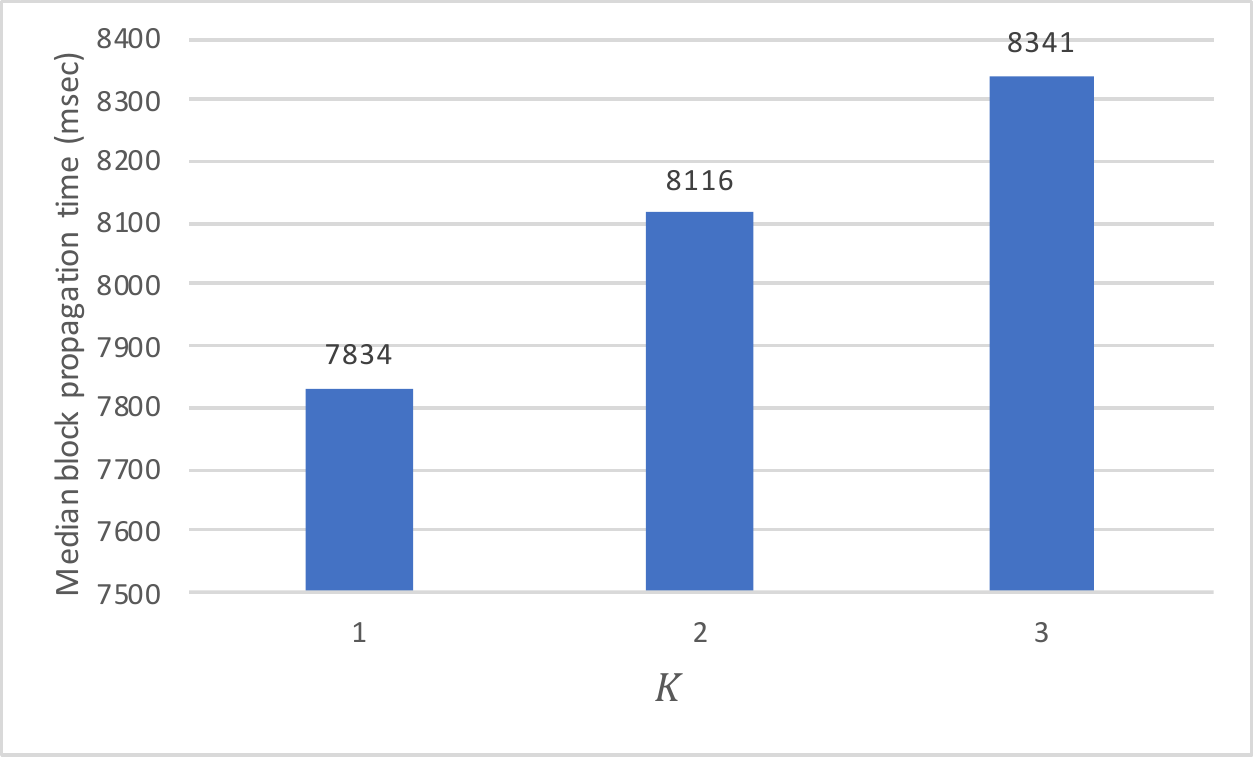} 
    \caption{Median block propagation time for different values of parameter $K$.}
    \label{fig:kresult}
  \end{center}
\vspace{-1.5ex}
\end{figure}

As in the parameter $P$, we performed the simulation until 5,000 blocks were generated; then, we measured the median value of the block propagation time and compared the average value of all 5,000 blocks.
In this experiment, $P =0.3$
Fig. \ref{fig:kresult} presents the block propagation time for different values of parameter K. It can be seen that the block propagation time is shortest for $K = 1$.

\begin{figure}[t]
  \begin{center}
    \includegraphics[width=8.35cm]{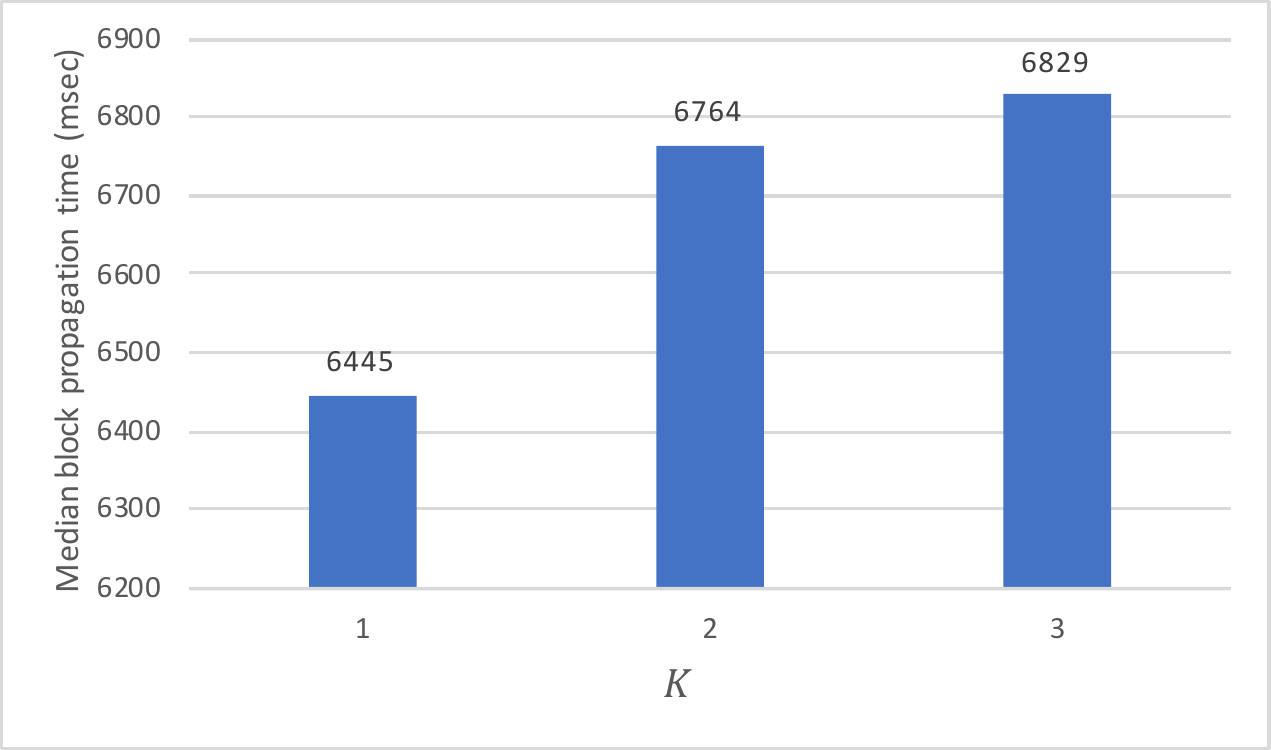} 
    \caption{Median block propagation time for different values of parameter $K$ (1,000 nodes).}
    \label{fig:kresult1000}
  \end{center}
\vspace{-3ex}
\end{figure}

It can be seen that the block propagation time is the shortest at $K=1$.

Next, we confirmed that the optimal parameters did not change depending on the number of nodes participating in the network.
Fig. \ref{fig:kresult1000} presents the experimental results when the number of participating nodes was changed to 1,000.

1,000. The results confirm that the tendency was identical for 1,000 nodes and 6,000 nodes.

\subsubsection{Modifying both $P$ and $K$}
Table \ref{tab:pk} presents the average value of the median block propagation times measured while changing both $P$ and $K$.

\begin{table}[t]
  \begin{center}
  \caption{Median block propagation time (ms) for parameters $P$ and $K$.}
    \label{tab:pk}
    \begin{tabular}{|l||c|c|c|c|c|} \hline
      \diagbox{$K$}{$P$} & 0.1 & 0.2 & 0.3 & 0.4 & 0.5  \\ \hline \hline
      1 & 7954 & 7835 & \textbf{7834} & 7845 & 8105\\\hline 
      2 & 8085 & 8117 & 8116 & 8120 & 8252    \\\hline 
      3 & 8159 & 8290 & 8341 & 8315 & 8296    \\ \hline
    \end{tabular}
        \end{center}
\end{table}

It can be seen that the propagation time is shortest when $K = 1$, regardless of the value of $P$.
In subsequent experiments in this study, we use $P = 0.3$ and $K = 1$, as they produce the shortest propagation time.

\subsection{Evaluation of proposed algorithm}
\begin{figure}[t]
 \begin{center}
    \includegraphics[width=8.35cm]{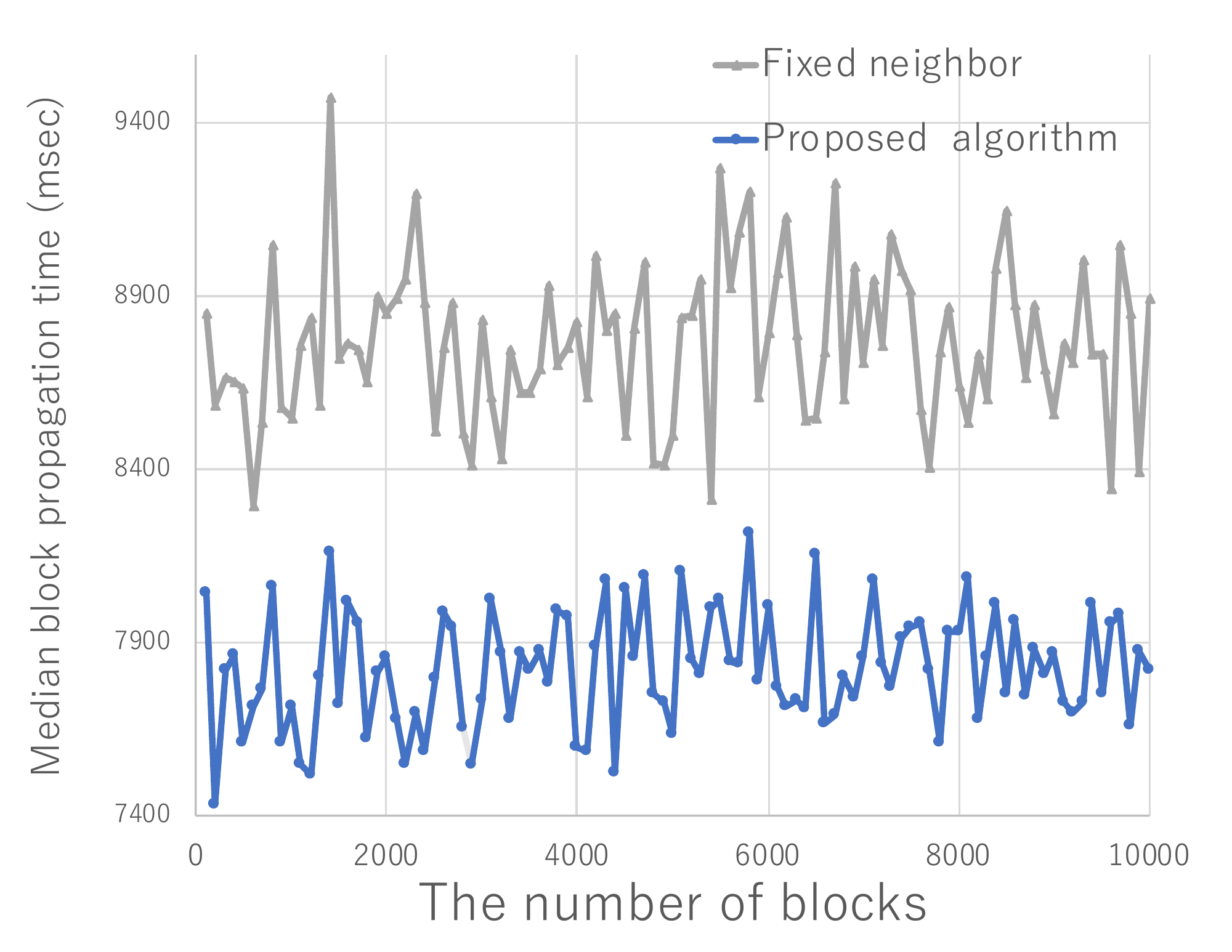} 
    \caption{Comparison of median block propagation times for a fixed-neighbor network and a network using the proposed neighbor selection algorithm.
     The horizontal axis denotes the number of generated blocks and the vertical axis indicates the median of the block propagation time.
The median of the block propagation time is averaged every 100 blocks and then plotted.
}
    \label{fig:mainresult}
  \end{center}
\vspace{-3ex}
\end{figure}

We compared a network with fixed neighbor nodes to a network with the proposed neighbor selection algorithm in terms of their median block propagation times.
The former network reproduces a regular Bitcoin network.
Fig. \ref{fig:mainresult} illustrates the comparison.

\begin{figure}[t]
 \begin{center}
    \includegraphics[width=8.35cm]{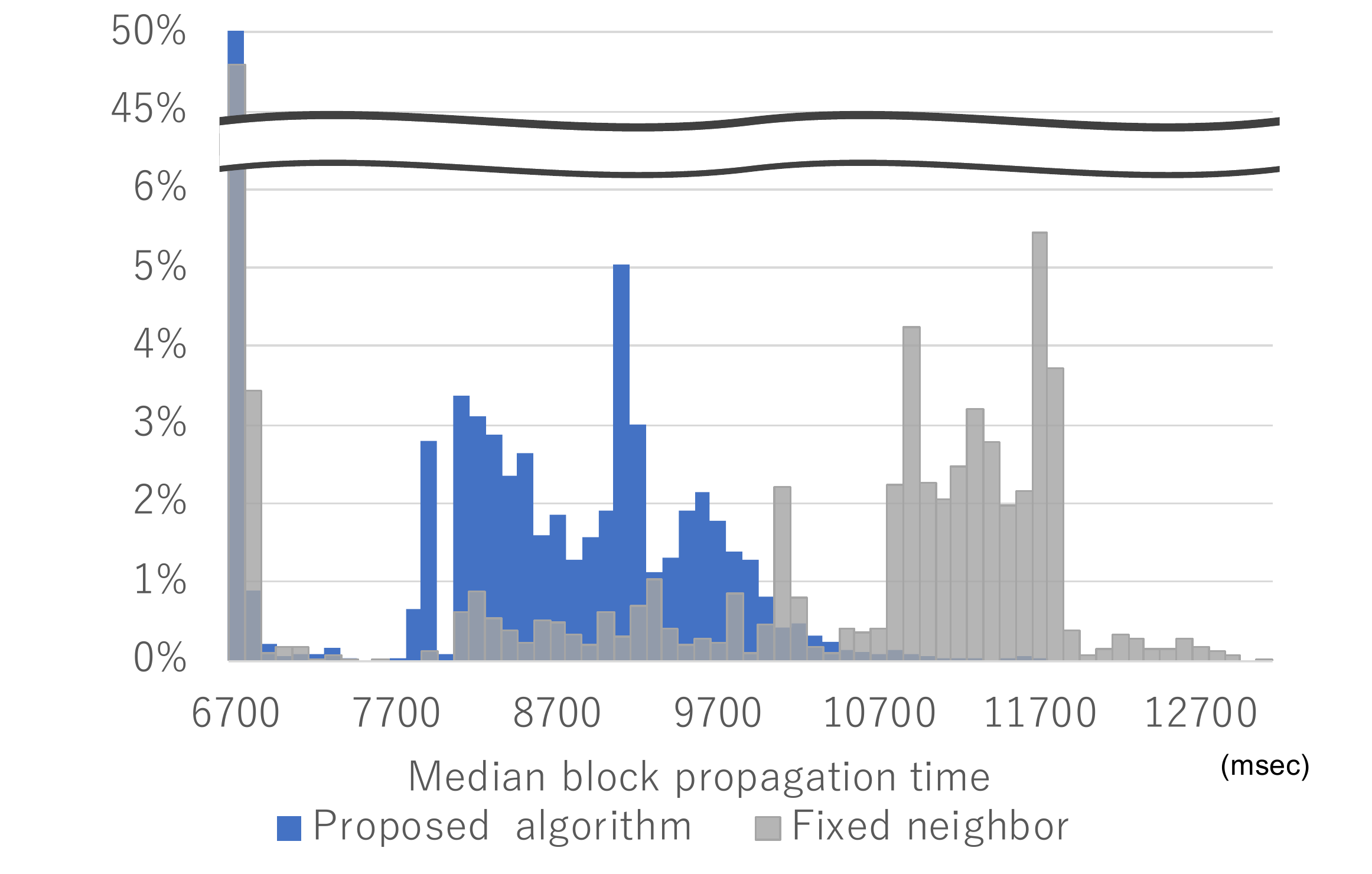} 
    \caption{Comparison of median propagation times for proposed algorithm and fixed-neighbor algorithm.}
    \label{fig:mainresult2}
  \end{center}
\vspace{-1.5ex}
\end{figure}

The experimental results reveal that the block propagation time is improved using the proposed neighbor node selection algorithm.
Additionally, the propagation time is sufficiently improved by replacing a small number of neighbor nodes.
After it is improved in the first several neighbor node updates, the propagation time exhibits no further changes. 

Fig. \ref{fig:mainresult2} presents a comparison of the
median propagation times between the proposed algorithm and the fixed-neighbor algorithm.

It can be seen that many propagation times are distributed between 6,700 ms to 6,900 ms, which at to the lower end of the time range, and that the proposed algorithm has a greater distribution near the lowest value than the fixed-neighbor algorithm does.
This is likely the case because the proposed algorithm allows more nodes to be concentrated in the central part of the network topology.
Additionally, for the proposed algorithm, a concentrated distribution also appears in the range of approximately 7,900 to 10,000 ms, while for the fixed-neighbor algorithm, a concentrated distribution appears in the range of approximately 10,200 to 12,000 ms.
The ranges of concentrated distribution reflect the propagation time of blocks generated at nodes far from the center of the network.
It can be seen that the proposed algorithm reduces the propagation time in this case.

\begin{figure}[t]
 \begin{center}
    \includegraphics[width=8.35cm]{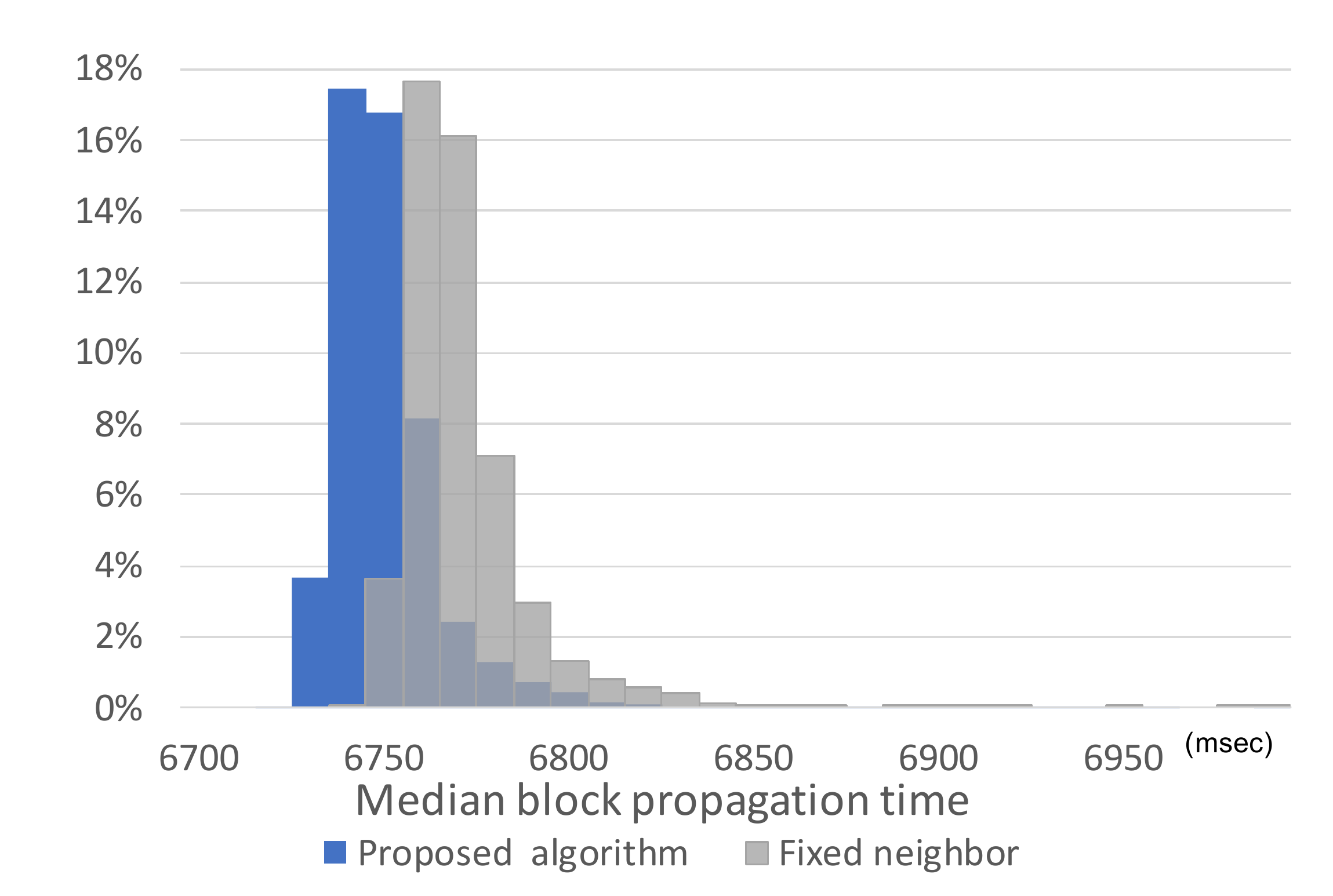} 
    \caption{Comparison of median propagation times (6,700 $\sim$ 7,000 ms) for proposed algorithm and fixed-neighbor algorithm.}
    \label{fig:mainresult3}
  \end{center}
\vspace{-3ex}
\end{figure}

Fig. \ref{fig:mainresult3} presents a comparison between algorithms for the range of 6,700 to 7,000 ms.

It can be seen that the proposed algorithm results in lower propagation times than the fixed-neighbor algorithm.
This indicates that with the proposed method, the propagation time of blocks generated at nodes near the center of the network is also improved.

\subsection{Influencing factors}
The proposed algorithm selects appropriate neighbor nodes using only the local information that each node can obtain, and the topology of the network changes.
Although the speed of block delivery is used as an indicator of neighbor node selection, it can be affected by both the network environment and block generation performance of nodes.
A node with a favorable network environment has a short waiting time for receiving blocks from other nodes and a short time for transmitting to other nodes; thus, the block delivery speed tends to be highter.
Nodes with high block generation performance have a high probability of generating blocks themselves. The blocks that they generate can be delivered earlier than any other node.
Below, we investigate how the block propagation time is affected by the network environment and block generation performance respectively.

\subsubsection{Effect of network environment}
\begin{figure}[t]
 \begin{center}
    \includegraphics[width=8.35cm]{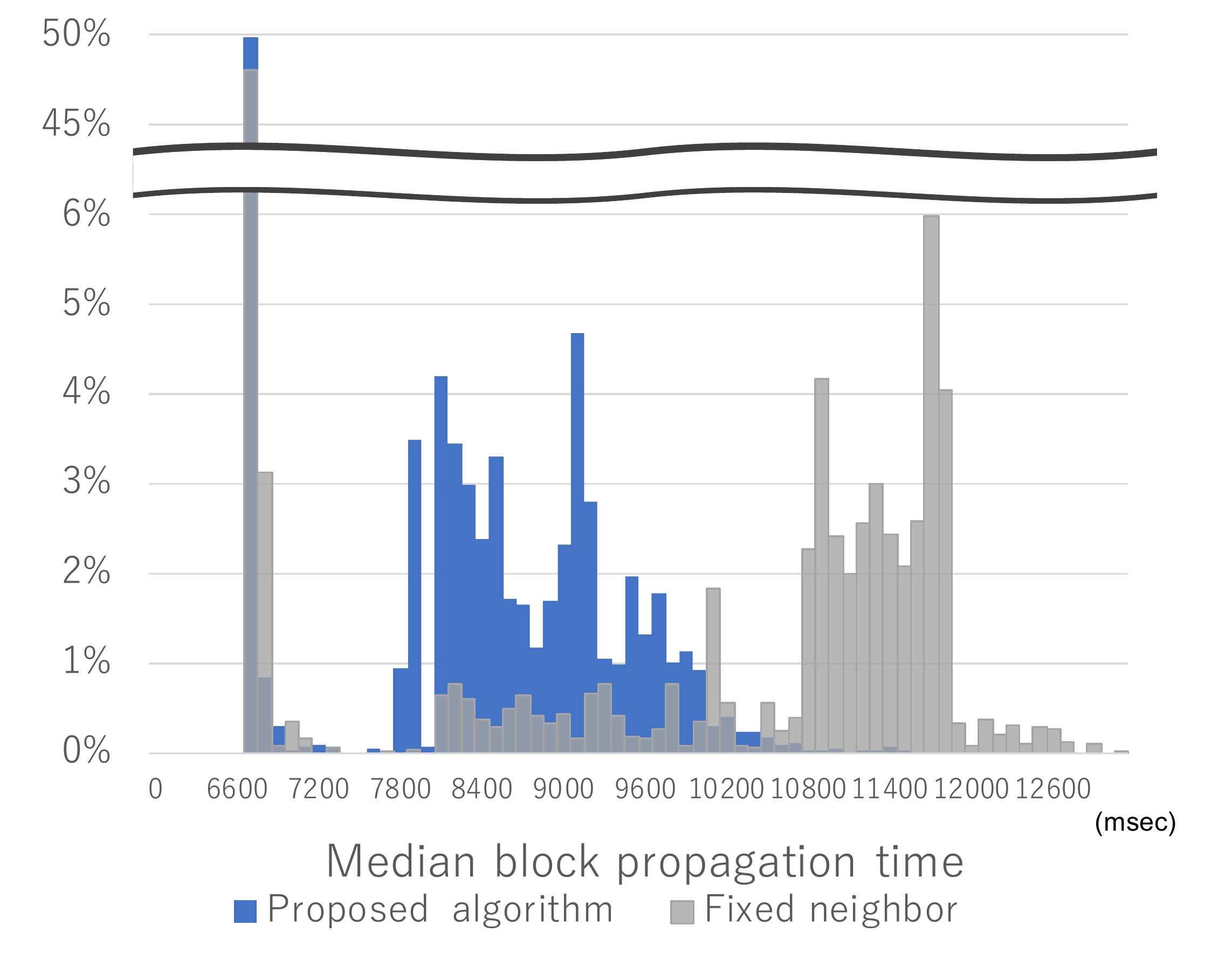} 
    \caption{Median of the propagation time for uniform block generation performance.}
    \label{fig:tiiki2}
  \end{center}
\vspace{-1.5ex}
\end{figure}

\begin{figure}[t]
 \begin{center}
    \includegraphics[width=8.35cm]{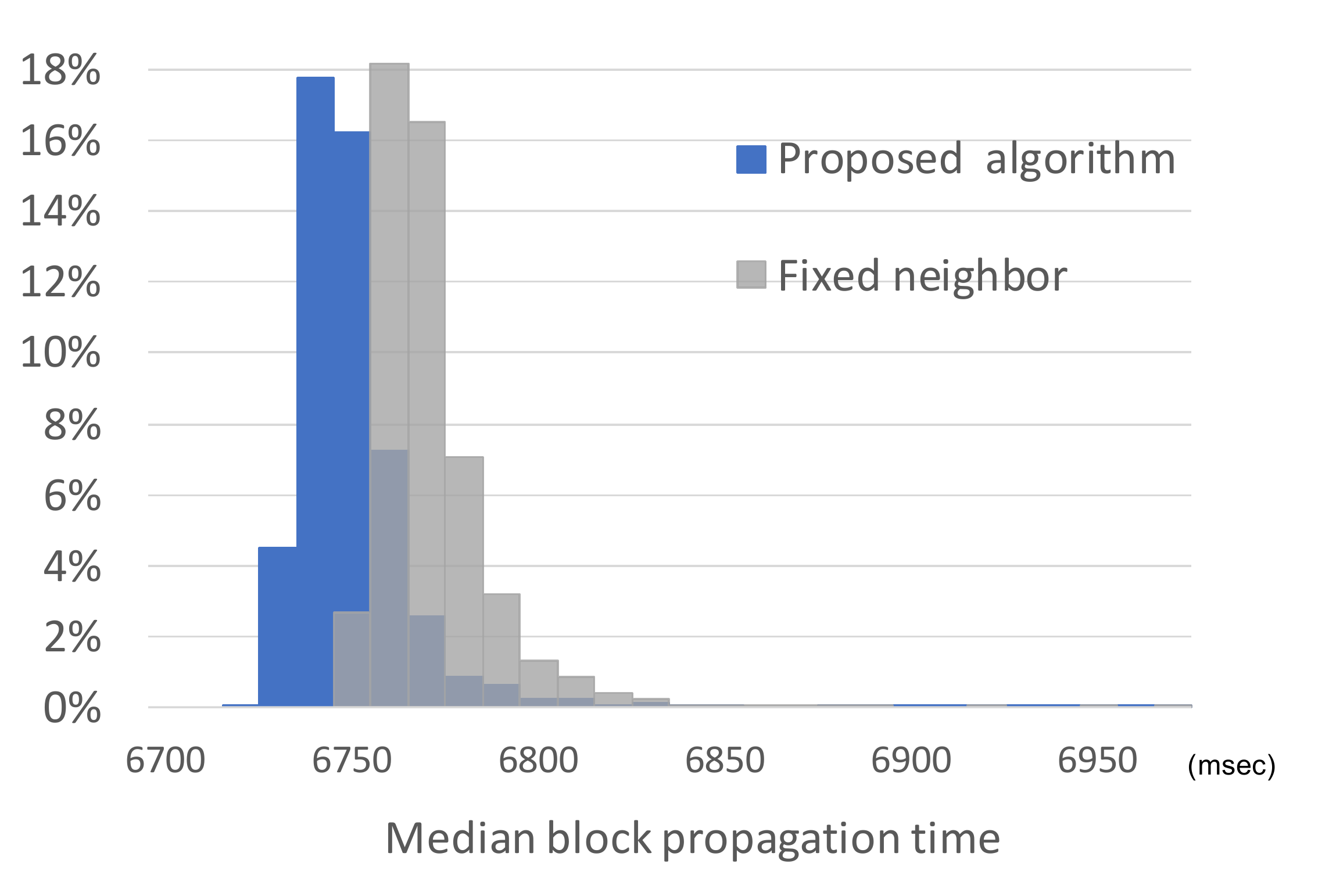} 
    \caption{Median of the propagation time for uniform block generation performance (6700 $\sim$ 7000 ms).}
    \label{fig:tiiki3}
  \end{center}
\vspace{-3ex}
\end{figure}

Here, we investigate the effect of the network environment of each node on the block propagation time for the proposed algorithm.
We eliminated the effect of block generation performance by setting the block generation performance of each node to a uniform value.
The other environmental conditions were the same as in the previous experiment.
The experiment was performed until 5,000 blocks were generated, and the median block propagation time was measured.
Fig. \ref{fig:tiiki2} presents a comparison of the median propagation times for the proposed algorithm and the fixed-neighbor algorithm, and Fig. \ref{fig:tiiki3} presents a comparison in the range of 6,700 to 7,000 ms.

These experimental result is similar to the previous one in Fig. \ref{fig:mainresult2} and Fig. \ref{fig:mainresult3}.
It can be concluded that network environment has a large impact on improving the block propagation time of the proposed algorithm.

\subsubsection{Effect of block generation performance}
\begin{figure}[t]
 \begin{center}
    \includegraphics[width=8.35cm]{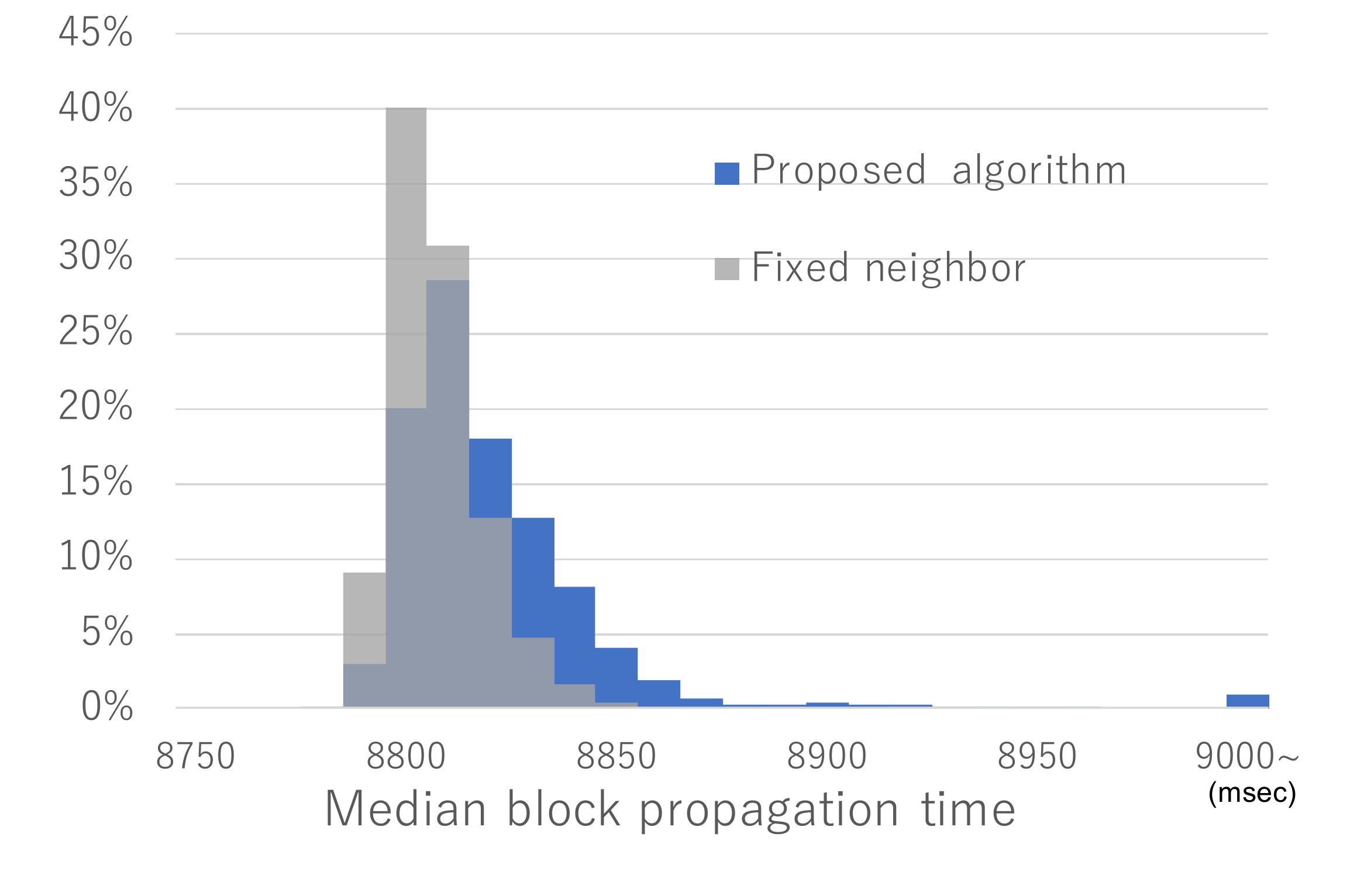} 
    \caption{Median of the propagation time for uniform network enviroment.}
    \label{fig:power2}
  \end{center}
\vspace{-3ex}
\end{figure}

Here, we investigate the effect of the block generation performance of each node on improving the block propagation time for the proposed algorithm.
The effect of network environment was eliminated by setting the propagation delay and bandwidth of each node to a uniform value.
The other environmental conditions were the same as in the previous experiment.
The experiment was performed until 5,000 blocks were generated, and the median block propagation time was measured.
Fig. \ref{fig:power2} presents a comparison of the median propagation times for the proposed algorithm and the fixed-neighbor algorithm.

It can be seen that the block propagation times are distributed in a very narrow range for a uniform network, and there is no improvement in propagation efficiency with the proposed algorithm.
In this experiment, the median propagation time of the top 100 nodes with the highest block generation performance was slightly lower than the median overall propagation time.
Nodes with high block generation performance are considered to be concentrated at the center of the network.
If nodes with a favorable network environment participate in a blockchain network, efficient block propagation may occur because nodes with high block delivery speed tend to be located near the center of network, where nodes with high block generation performance are concentrated. 

\section{Conclusion}
\label{sec:summary}
The major drawbacks of blockchains include low transaction throughput and the long approval times.
These problems can be addressed by shortening the block generation interval; however, if the block generation interval alone is shortened, the occurrence of forks rises and the security risk increases.
Thus, it is necessary to shorten the block propagation time in order to shorten the block generation interval while suppressing the occurrence of forks.
In this study, we developed a neighbor node selection algorithm to form a network with a short block propagation time.
In a blockchain network, each node rates other nodes according to the speed of block delivery, and preferentially selects a node with a good score to be a neighbor node.
With our proposed algorithm, a network with hight block propagation efficiency can be formed using only information that each node can obtain locally.
Using a simulator, we confirmed that the proposed neighbor node selection algorithm improves block propagation time.

Future work should examine additional block propagation protocols used in existing blockchains.
Protocols other than the traditional ones used in this paper have been proposed.
A famous example is the compact block relay \cite{bip-0152}.
In a traditional block propagation protocol, if the transaction in a received block is already present in the transaction pool, this implies that the same transaction has been received twice．
In the compact block relay protocol, when receiving block, the bandwidth used in suppressed by receiving only transactions that are not included in the transaction pool.
When the proposed algorithm is used in a blockchain employing compact relays, the influence on bandwidth of neighbor node selection is reduced, and the propagation delay and block generation performance are expected to improve.

Future work should also investigate changes to the security of the network resulting from the proposed neighbor node selection algorithm.
Because the proposed algorithm alters the network topology, we must investigate what influence it has on resistance to fragmentation and other attributes by analyzing the graph of the network.

\section*{Acknowledgment}
This work was supported by
SECOM Science and Technology Foundation.

\bibliography{bib}
\bibliographystyle{unsrt}

\vspace{12pt}

\end{document}